\documentstyle[12pt]{article}
\newcommand{\newc}{\newcommand} 
\newc{\ra}{\rightarrow} 
\newc{\lra}{\leftrightarrow} 
\newc{\beq}{\begin{equation}} 
\newc{\eeq}{\end{equation}} 
\newc{\barr}{\begin{eqnarray}} 
\newc{\earr}{\end{eqnarray}} 

\def\mue{(\mu ^-,e^-)}
\def\al{\alpha}
\def\be{\beta}
\def\del{\delta}
\def\la{\lambda}

 1

\begin{document} 
\begin{titlepage}

\begin{center}
{\Large \bf State-by-state calculations for all channels of \\
the exotic $\mue$ conversion process}

\vspace{12mm}

T. S.   KOSMAS \footnote{Permanent address: Theoretical Physics Division,
University of Ioannina, GR 451 10, Greece.}, 
AMAND  FAESSLER, F. \v SIMKOVIC,

\vspace{2mm}

{\it Institut f\"ur Theoretische Physik der Universit\"at
T\"ubingen, D-72076 T\"ubingen, Germany}

\vspace{5mm}

J. D. VERGADOS

\vspace{2mm}

{ \it Theoretical Physics Division, University of Ioannina,
GR-451 10 Ioannina, Greece }
\end{center}

\vspace{10mm}
\begin{abstract}
{\it The coherent and incoherent channels of the neutrinoless muon to electron
conversion in nuclei, $\mu^- \,(A,Z) \ra e^- \,(A,Z)^*$, are studied 
throughout the periodic table. The relevant nuclear matrix elements are 
computed by explicitly constructing all possible final nuclear states in the 
context of the  quasi-particle RPA.
The obtained results are discussed in view of the existing at PSI and TRIUMF
experimental data for $^{48}Ti$ and $^{208}Pb$ and compared with results
obtained by: (i) shell model sum-rule techniques (ii) nuclear matter mapped 
into nuclei via a local density approximation and (iii) earlier similar 
calculations.}
\end{abstract}

\vspace{1.7cm}

PACS number(s): 25.30.-c, 23.40.Bw, 13.35.Bv, 21.60.Jz, 12.60.Cn

\vspace{3.0cm}

\end{titlepage}

\centerline{\bf I.  INTRODUCTION}
\bigskip

The neutrinoless muon to electron conversion in the field of a nucleus,

\beq
\mu ^- \,\, + \, \, (A,Z)\, \ra \,e^-\,\, + \,\, (A,Z)^*
\label{eq:I.1}
\eeq

\noindent
is forbidden in the standard model by lepton flavour conservation and plays an 
important role in the study of flavour changing neutral
currents which violate muon and electron numbers \cite{DOHM}-\cite{KLV94}.
Within the last decade, experiments at PSI \cite{DOHM,SINDR,Honec} and TRIUMF
\cite{Ahmad} aiming at a search of $\mu -e$ conversion electrons have not
yet observed such events. These experiments have, however, provided us with
useful constraints for the violation of muon and electron numbers. The best
upper limit on the branching ratio

\beq
R^{Ti}_{\mu e} =  {\Gamma(\mu ^-,e^-)} / {\Gamma(\mu^-,\nu_{\mu})} \,
 < \, 4.3\times 10^{-12}
\label{eq:I.2}
\eeq

\noindent
has recently been set by SINDRUM II at PSI \cite{DOHM} by using $^{48}Ti$ as 
target. This value is of the same order with the previous limit set at TRIUMF
\cite{Ahmad}, i.e. $R^{Ti}_{\mu e}<4.6\times 10^{-12}$. This year \cite{Honec},
experimental data extracted at PSI by using $^{208}Pb$ have yielded an upper 
limit $R^{Pb}_{\mu e} < 4.6 \times 10^{-11}$.
This experiment improved by an order of magnitude over the previous upper 
limit, $ R^{Pb}_{\mu e} < 4.9\times 10^{-10}, $ extracted
from preliminary experimental data for the same target
at TRIUMF \cite{Ahmad}. 

The experimental sensitivity is expected to be further improved by two to three
orders of magnitude by on going experiments at PSI (to $10^{-14}$) \cite{DOHM}, 
at TRIUMF (to $10^{-14}$) \cite{Ahmad,Depom} and at INS (to $10^{-14} -
10^{-16}$) \cite{INS}. Hopefully, such experiments will not only yield a still 
better limit, but they will detect some $\mue$ events  which will signal the 
break down of the muon number conservation revealing "new physics" 
beyond the standard model. For a discussion of lepton flavour 
violation limits in conjunction with theoretical predictions, the reader is 
referred to the recent survey by Depommier and Leroy \cite{Depom}.

Due to the similarity of electrons and muons, $\mu -e$ conversion was 
originally expected to proceed real fast. From a theoretical point of view, 
the basic background for the $\mue$ has been set long time ago by Weinberg and
Feinberg \cite{WeiFei} who assumed that this process is mediated by virtual
photons (Fig. 1(a)). Non-photonic contributions (Fig. 1(b)-(d)) have been 
included later on in the post gauge theory era (for a recent review on this 
topic see Ref. \cite{KLV94} and for the experimental data extracted from 
various targets see Ref. \cite{Brym}.

One expects that the Z-exchange diagrams, Fig. 1(b),(c), to be less important
than the W-box diagrams, Fig. 1(d), even for the incoherent process. The precise
value depends, of course, on details like quark masses etc. One also expects
the W-box to dominate by large factors especially in the case of heavy 
intermediate neutrinos \cite{Leonta}. Similar conclusions have been obtained 
by Marciano and Sanda \cite{MarSan}. In the present work we have not included 
exotic particles like $Z^{\prime}$ \cite{Berna}, exotic Higgs scalars,
many Higgs doublets, R-parity violating interactions etc., which may in some
models be important. We intentionally stayed within the context of the minimal
extensions of the standard model keeping also in mind that our emphasis here is
on the nuclear structure aspects.

An interesting feature of the $\mue$ conversion in nuclei is the possibility of
the ground state to ground state transitions. The strength of this channel is
expected to be enhanced because of the coherent contribution of all nucleons
of the participating nucleus or at least all protons.
The rate for such transitions can be expressed in terms of the proton and 
neutron nuclear form factors \cite{Shank,VER,KV88}. Earlier estimates for the 
branching ratio $R_{\mu e}$ by Weinberg and Feinberg \cite{WeiFei} have 
indicated that, for $A \ge 100$ this ratio is approximately constant, while 
Shanker \cite{Shank} found that the ratio $R_{\mu e}$ could be bigger in heavy 
nuclei.

The incoherent rate is much harder to calculate. The first such calculations
have been performed only recently \cite{KV90,KV89} in nuclei with closed
(sub)shells  throughout the periodic table by employing shell model sum-rules,
i.e. by invoking closure approximation in some form  with a suitable choice of
a mean excitation energy, using a single Slater determinant for the initial
state. In reality, these calculations give the total rate. The incoherent
strength can be estimated by subtracting from the total strength the coherent
part obtained independently. What, however, is needed is the ratio of the
coherent rate to the total rate. Shell model results showed that the coherent
channel dominates the $\mue$ process for light and medium nuclei, but in the
region of $^{208}Pb$, a great part of the rate comes from the inelastic
channels. Furthermore, these calculations showed that the dependence of the
branching ratio $R_{\mu e}$ on the nuclear mass  $A$ and charge $Z$ reaches
a maximum around $A \sim 100$ in agreement with the estimates of Ref.
\cite{WeiFei}.

Recently \cite{Chia}, we have employed, for the coherent and incoherent $\mue$ 
conversion, another approach based on a local density approximation
in conjunction with a relativistic Lindhard function for the description of
the elementary processes: $\mu^-p \ra e^-p$ and $\mu^-n \ra e^-n$.
The incoherent rate in this method was obtained by integrating
over the excited states of a local Fermi sea. 
These results have shown that, the coherent contribution is dominant for
all nuclei and that the branching ratio $R_{\mu e}$ presents a maximum in the
region of very heavy nuclei, i.e. in the $Pb$ region.

In yet another recent theoretical study of the $\mue$ conversion \cite{KVCF},
the quasi-particle RPA (QRPA) was employed for the construction of the
final nuclear states entering the coherent and incoherent rate. Such
quasi-particle RPA results for $^{48}Ti$ have shown that the coherent channel 
dominates. One of the advantages of the QRPA method is that it can be
used to estimate the mean excitation energy of the nucleus of 
interest which in turn is useful in checking the results 
of the above mentioned closure approximation which are
sensitive to this property. The important result \cite{KVCF} was that the mean
excitation energy $\bar E$ of the nucleus in process (1) is very small, 
$\bar E \approx 1 MeV$, which is appreciably smaller than  $\bar E
\approx 20 MeV$ used in shell model calculations \cite{KV90}. The latter value
had been chosen from the phenomenology of the charge changing
($\mu^-,\nu_{\mu}$) reaction. This difference is mainly due to the fact that,
the coherent elastic channel, possible only in the $\mue$, is the
dominant channel.

From the above discussion it is clear that, a detailed study of all possible
channels of the $\mue$ conversion for medium and heavy nuclei and in particular
for nuclei around $^{208}Pb$, which is of current experimental interest, is 
needed. In the present work, we use the formalism developed in the context of
quasi-particle RPA \cite{KVCF} (improved in the part of the reduced matrix 
elements, see appendix) to extend our previous results for $^{48}Ti$ and we 
report calculations performed for all individual $\mue$ conversion channels, 
in a set of isotopes covering the above region (see below Table I) including, 
of course, $^{48}Ti$ and $^{208}Pb$, since, the upper limit on the
branching ratio $R_{\mu e}$ has been extracted \cite{DOHM}-\cite{Honec} 
from experimental data on these nuclei. We note that, the method using a local
density approximation \cite{Chia} cannot give us the individual contribution 
of each accessible channel.

Before embarking on such calculations, we mention that for certain nuclei, in
particular those with closed shells, like $^{60}Ni$ and $^{208}Pb$, a
special treatment in QRPA is required in order to determine the pairing
parameters for protons ($g^p_{pair}$) and neutrons ($g^n_{pair}$). In this
work we follow the manner used recently in the double beta decay
\cite{Suho93}. In sect. II we briefly discuss the method used, while the
obtained results are presented and discussed in sect. III and the conclusions
are summarized in sect. IV.

\bigskip
\bigskip
\centerline{\bf II.  BRIEF DESCRIPTION OF THE METHOD}
\bigskip

\centerline{\bf A. The $\mue$ conversion effective operator}
\bigskip

The effective Hamiltonian operator of the $\mue$ conversion, which involves 
both vector and axial vector currents, after the usual non-relativistic 
reduction takes the form \cite{KLV94}

\beq
\Omega_V = \tilde g_V{\sum^{A}_{j=1}}\Big (
 {3+f_V\be\tau_{3j}}\Big)
e^{-i{\bf q} \cdot {\bf r}_j} ,\quad \quad
{\bf \Omega}_A =-\tilde g_Af_A{\sum^{A}_{j=1}}\Big (
{\xi +\be\tau_{3j}}\Big) {{{\bf \sigma}_j}\over
\sqrt 3}\,e^{-i{\bf q} \cdot {\bf r}_j} 
\label{eq:IIA.1}
\eeq

\noindent
The parameters ${\tilde g}_V, \,\,\, {\tilde g}_A$ and $ \beta$ depend on the
assumed mechanism for lepton flavour violation \cite{VER,KLV94}. For the 
photonic mechanism these parameters take the values ${\tilde g}_V=1/6$,
${\tilde g}_A = 0$, $\beta =3$, $f_V=1$ while, for the non-photonic neutrino
mediated mechanism, they are ${\tilde g}_V={\tilde g}_A= 1/2$, $\beta
=5/6 $, $f_V=1$, $f_A=1.24$ and $\xi= f_V/f_A = 1/1.24$. 

In Eq. (\ref{eq:IIA.1}), ${\bf q}$ represents the momentum transfer
to the nucleus. Its magnitude is approximately given by

\beq
q \, \, = \, \mid {\bf q}\mid \, =\,\, m_\mu -\epsilon_b -(E_f-E_{gs})
\label{eq:IIA.2}
\eeq

\noindent
where $m_{\mu}$ is the muon mass, $\epsilon_b$ is the muon binding energy
and $E_f$, $E_{gs}$ are the energies of the final and ground state of the
nucleus, respectively. We should mention that $\epsilon_b$, although
negligible in light nuclei, can become important in heavy elements (see below
Table I).

The matrix elements of the operators of Eq. (\ref{eq:IIA.1}) can be obtained 
via the multipole operators $T_M^{(l,s)J}$ given in the appendix (for details 
see Refs. \cite{KLV94,KVCF}). 

\bigskip
\centerline{\bf B. Nuclear matrix elements for the coherent rate}
\bigskip

In the case of the coherent $\mue$  process, i.e. ground state to ground state
transitions ($0^+\ra 0^+$), only the vector component of the $\mue$ operator
contributes and the coherent rate is proportional to \cite{KLV94}
\beq
\mid <f \mid \Omega (q)\mid i,\mu> \mid ^2
=  \tilde g^2_V \,(3 + f_V \beta)^2 \, \Big[
 \, \tilde F_{p} (q^2) + \frac{3-f_V\beta}{3+f_V\beta}
\tilde F_{n} (q^2) \Big]^2
\label{eq:IIB.1}
\eeq

\noindent
where 

\beq
\tilde F_{p,n} (q^2) = \int d^{3}x \; \rho_{p,n} ({\bf x}) \; e^{-
i{\bf q} \cdot {\bf x}}\; \Phi_{\mu} ({\bf x})
\label{eq:IIB.2}
\eeq

\noindent
In the last equation $\rho_{p}({\bf x})$, $\rho_{n}({\bf x})$ represent
the proton, neutron densities normalized to Z and N, respectively
and $\Phi_{\mu} ({\bf x})$ is the muon wave function. If we assume that the
muon is bound in the 1s atomic orbit which
varies very little inside the nucleus, we can factorize the muon wave function
out of the integral of Eq. (\ref{eq:IIB.2}) and write

\beq
\tilde F_p (q^2) \approx  < \Phi_{1s}> \, ZF_Z(q^2), \qquad
\tilde F_n (q^2) \approx  < \Phi_{1s}> \, NF_N(q^2)
\label{eq:IIB.3}
\eeq

\noindent
where ${F}_{Z}$  $({F}_{N}$) the usual proton (neutron) nuclear form factors. 

We should mention that, experimentally the most interesting quantities are the
branching ratio $R_{\mu e}$ and the ratio $\eta$ of the coherent to the total
$\mue$ conversion rate (see Eq. (\ref{eq:IIIC.1}) below). We do not expect 
the branching ratio to be greatly affected by the approximation of Eq. 
(\ref{eq:IIB.3}), especially if we calculate the total $\mu^-$ capture rate in 
the same way. We expect this to be good even if for the total muon capture 
rate we use the Primakoff function \cite{Goul-Prim}, which is obtained by 
explicitly using this approximation. The Primakoff function fits the 
experimental data remarkably well throughout the periodic table (even for heavy 
nuclei). Furthermore, and for similar reasons, we expect that the ratio 
$\eta$ is not going to be drastically affected by this approximation. 

In the above approximation the nuclear dependence of the rate for the coherent
process, is proportional to the matrix element

\beq
M^2_{coh}(q^2) \, \equiv  \, 
M^2_{gs \ra gs}(q^2) =   \, 
 \left [1+\frac{3-f_V\beta}{3+f_V\beta}\,
\frac{N}{Z} \, \frac{F_N(q^2)}{F_Z(q^2)}\right ]^2 \,
Z^2F_Z^2(q^2)
\label{eq:IIB.4}
\eeq

Thus, the variation of the coherent $\mue$ conversion rate through the periodic
table can be studied by calculating the matrix elements $M^2_{gs \ra gs}$ of
Eq. (\ref{eq:IIB.4}) for various A and Z. 
The nuclear form factors involved in $M^2_{gs \ra gs}$ can either  be calculated
by using various models as shell model \cite{KV88,KV92}, quasi-particle
RPA \cite{KVCF,KFSV94} etc., or can be obtained directly from experiment
whenever possible \cite{Heisen,Vries}.

In the context of the quasi-particle RPA with an uncorrelated vacuum as ground
state, the nuclear form factors are given by (see appendix)

\beq
 F_{\tau} (q^2)\,=
\,{1 \over \tau} \,{\sum_{j}}\,\, \Big( V^{\tau}_j\Big)^2 \, (2 j+1)\, 
<j  \mid j_0(q r) \mid j >, \,  \qquad \tau = Z, \,\, N
\label{eq:IIB.5}
\eeq

\noindent
where $\Big( V^{\tau}_j \Big)^2$ are the occupation probabilities for the
proton, neutron single particle states  $\mid j >$  included in the used model
space ($j \,\equiv \,(n,l,j) $).

We should mention that, in the photonic case ($\beta =3$) only the protons
of the considered nucleus contribute and the right hand side of
Eq. (\ref{eq:IIB.4}) becomes $Z^2F_Z^2(q^2)$.

\bigskip
\centerline{\bf C. Incoherent rate by explicit calculations of the final states}
\bigskip

The incoherent $\mue$ conversion rate is evaluated by summing the partial
rates for all final nuclear states $\mid f>$ except the ground state. We
need calculate the matrix elements for both the vector and axial vector
operators of Eq. (\ref{eq:IIA.1}), i.e. the quantities

\beq
S_{\al} = {\sum_f}
\Big({ {q_f}\over {m_{\mu}}}\Big)^2\int { { d {\hat {\bf q}}_f}
\over {4 \pi}} {\mid <f \mid {\Omega}_{\al} \mid gs > \mid}^2,
\qquad \mid f > \ne \mid gs >,
\qquad \al=V,A.
\label{eq:IIC.1}
\eeq

\noindent
(${\bf {\hat q}}_f$ is the unit vector in the direction of the momentum
transfer ${\bf q}_f$). 

As we have mentioned in the introduction, for the calculation of $S_V$ and 
$S_{A}$, one can either use closure approximation (in which case the state 
$\mid f > = \mid gs > $ is
included in Eq. (\ref{eq:IIC.1})) or compute state-by-state the partial rates
involved if one can construct the final states $\mid f>$ in the context of some
nuclear model. By using the multipole expansion operators 
${\hat T}^{(l,\sigma)J}$ (see appendix), the matrix elements $S_{V}$ and 
$S_{A}$ are written as

\beq
 S_{\al} =  \sum_{f_{exc}}  \Big( {{q_{exc}} \over {{m_{\mu}}}} 
\Big)^2 {\sum_{l,J}}\, {\mid <f_{exc} \mid
\mid {\hat T}^{(l,\sigma)J}\mid \mid gs >\mid}^2
\label{eq:IIC.2}
\eeq

\noindent
($\al \, = \,V,\, A$, for the vector ($\sigma=0$), axial vector ($\sigma=1$) 
component, respectively). The partial matrix element from the initial state 
$0^+$ to an excited state $\mid f>$ in the context of QRPA takes the form
 
\beq
 <f \mid \mid {\hat T}^{(l,\sigma)J}\mid \mid 0^+ > = {\sum_{\la,\tau}} \,\,
W_{\la}^J\,
 \Big [\,\, X_{\la}^{(f,J,\tau)} U^{(\tau)}_{j_2}V^{(\tau)}_{j_1}  +
 Y_{\la}^{(f,J,\tau)} V^{(\tau)}_{j_2}U^{(\tau)}_{j_1}
  \,\, \Big]
\label{eq:IIC.3}
\eeq

\noindent
where $V^{(\tau)}_j$ and $U^{(\tau)}_j$ represent the probability amplitudes 
for the single particle states to be occupied and unoccupied, respectively.
They are determined by solving the BCS equations iteratively. X and Y represent
the forward and backward scattering amplitudes. They are obtained by solving 
the QRPA equations. The index $\la$ runs over two particle configurations 
coupled to a given J, namely $(j_1,j_2)J$ for the proton ($\tau=1$) or neutron 
($\tau=-1$). The quantities $W_{\la}^J \equiv W_{j_2j_1}^J$ are given in the 
appendix.

For the total $\mue$ rate the relevant matrix elements are obtained by adding
the vector and axial vector contributions of the coherent and incoherent rate
i.e.

\beq
 M^2_{tot} = S_V + 3 S_A + S_0
\label{eq:IIC.4}
\eeq

\noindent
where $S_0$ is associated with the ground state to ground state transition

\beq
 S_0 = \Big({{ q_{gs}} \over {m_{\mu}}}\Big)^2 
{\sum_{l,J}}\, {\mid <gs
\mid \mid {\hat T}^{(l,\sigma)J}\mid \mid gs > \mid}^2 
\label{eq:IIC.5}
\eeq

\noindent
for the vector component $\sigma=0$, and for the axial vector component
$\sigma=1$. For $0^+$ nuclei only the vector term contributes.

\bigskip
\bigskip
\centerline{\bf III. RESULTS AND DISCUSSION}
\bigskip

Using the method outlined above, in the present work we have calculated the 
matrix elements needed for both the coherent and incoherent $\mue$ rate, for 
the nuclei $^{48}Ti$, $^{60}Ni$, $^{72}Ge$, $^{112}Cd$, $^{162}Yb$ and 
$^{208}Pb$. The specific parameters used and a brief description of the model 
spaces employed can be read from Table I. For all nuclei considered we have 
employed the same model space for protons and neutrons. For the harmonic
oscillator parameter $b$ in the region of heavy nuclei we have employed the
improved expressions of Ref. \cite{Lala}.

In the BCS description of the uncorrelated ground state, for each nuclear
isotope the single particle energies have been calculated from a Coulomb
corrected Wood-Saxon potential with spin-orbit coupling. The G-matrix 
elements of the realistic Bonn one-boson exchange potential \cite{Holin}
have been employed. The values of pairing parameters, $g^p_{pair}$ and
$g^n_{pair}$, renormalizing the proton and neutron pairing channels
in the G-matrix have been deduced by comparing the quasi-particle energies
with experimental pairing gaps as is described in Refs. \cite{Suho88,CheSim}. 
For the special cases of $^{60}_{28}Ni$, which is a
proton closed-shell nucleus and $^{208}_{82}Pb$, which is a doubly
closed-shell nucleus, the pairing parameters have been deduced from the 
neighboring nuclei $^{60}_{26}Fe$ and $^{208}_{84}Po$, respectively, in
analogy with the procedure followed in the study of the nuclear double beta
decay  in the double closed shell nucleus $^{48}_{20}Ca$ \cite{Suho93}. The
resulting  pairing parameters $g^p_{pair}$, $g^n_{pair}$ for each nucleus  are
shown in Table I.

\bigskip
\centerline{\bf A. The coherent process}
\bigskip

It is obvious from Eq. (\ref{eq:IIB.4}) that, for the coherent process, 
i.e. $gs
\ra gs$ transitions, we need the proton and neutron  nuclear form factors,
$F_Z(q^2)$ and $F_N(q^2)$, respectively. The results obtained by using as ground
state the uncorrelated RPA vacuum, are listed in Table II for the
following two cases: 

(i) By neglecting the muon binding energy $\epsilon_b$ in Eq. (\ref{eq:IIA.2}) 
(as in Refs. \cite{WeiFei,KV90}). 
Then, the elastic momentum transfer is the same for all nuclei, i.e. $q
\approx m_{\mu} \approx .535 fm^{-1}$. Such results  are indicated as QRPA(i).

(ii) By taking into account $\epsilon_b$ in Eq. (\ref{eq:IIA.2}). Then, the 
elastic momentum transfer is  $q \approx m_{\mu} - \epsilon _b$ and
varies from $q \approx .529 fm^{-1}$, for $^{48}Ti$ where $\epsilon_b
\approx 1.3 MeV$, to  $q \approx .482 fm^{-1}$, for $^{208}Pb$ where
$\epsilon_b \approx 10.5 MeV$ (see Table II, results indicated as QRPA(ii)).

In Table II we also present the shell model results of Ref. \cite{KV90},
obtained with $q =.535 fm^{-1}$ throughout the periodic table, i.e. as
in case (i) above. We see that, QRPA(i) and shell model methods,
give about the same results. However, the form factors of QRPA(ii) for heavy
nuclei differ appreciably from those of both QRPA(i) and shell model.
For $^{208}Pb$, for example, the QRPA(ii) form factors are about $30\%$
larger than the corresponding QRPA(i) and shell model.
This happens because the inclusion of $\epsilon_b$ results in a smaller momentum
transfer to the nucleus and, consequently, in an increase of the form factors.
The larger the value of $\epsilon_b$, lead region, the bigger the difference
between form factors QRPA(i) and QRPA(ii). 

In Table II we also show the experimental form factors obtained from electron 
scattering data \cite{Heisen,Vries} at momentum transfer $q = m_{\mu}
-\epsilon_b$. We see that, when using the right form factors, i.e. taking into 
account the binding energy $\epsilon_b$ (QRPA(ii) case), the form factors 
calculated in the present work, are in good agreement with the experimental 
ones. The deviation is less than $5\%$ with the possible exception of 
$^{112}Cd$ and $^{208}Pb$ where it is about $10\%$.  

The variation of the coherent nuclear matrix elements $M^2_{coh}$ with
respect to A and Z is shown in Fig. 2(a), for the photonic mechanism, and
Fig. 2(b), for the non-photonic one. From these Figures we see that, 
by taking into account the muon binding energy $\epsilon_b$,
QRPA(ii), all matrix elements increase continuously up to the lead region
where they become about a factor of two bigger than the corresponding QRPA(i)
and shell model values. This implies that the coherent rate becomes bigger for
heavy nuclei, $Pb$ region, which makes such nuclei attractive from an 
experimental point of view \cite{Ahmad,Honec} provided, of course, 
that they also satisfy other additional criteria, e.g.
the minimization of the reaction background etc. \cite{Depom,Brym}. The
$\mue$ conversion  electrons  of a given target are expected to show a
pronounced peak around $E_e = m_{\mu}- \epsilon _b $, which for the lead 
region is $E_e \approx 95 MeV.$  One prefers this peak to be as far as 
possible above the reaction induced background. 

We should recall that, in the present work and in shell model method of Ref.
\cite{KV90}, the factorization approximation Eq. (\ref{eq:IIB.3}) was used. 
The exact expression, Eq. (\ref{eq:IIB.2}), was used in Ref. \cite{Chia} and 
yielded matrix elements which for heavy nuclei are bigger than the approximate 
ones. This, however, as we have extensively seen in sect. II.B, only slightly 
affects the branching ratio $R_{\mu e}$ and the ratio $\eta$ of the coherent
rate to the total rate, which in our case are the most important quantities.  
We also mention that, shell model results for the total muon capture rate
\cite{Duplain}, obtained by using the exact muon wave function, differ by only 
5.7\%, in the case of $^{60}Ni$, and by 7.0\%, in the case of $^{208}Pb$, from 
those obtained by using the approximation of Eq. (\ref{eq:IIB.3}). Detailed 
QRPA calculations, which do not invoke this factorization approximation, are 
under way and will be published elsewhere. 

\bigskip
\centerline{\bf B. Incoherent process}
\bigskip

As we have stated in sect. II.C, the incoherent process in the present work is
investigated by calculating state-by-state the contributions of all the excited
states of the nucleus in question which are included in the model space
described in Table I.

For the photonic mechanism the nuclear matrix elements obtained for all positive
and negative parity states up to $6^-$, $6^+$ are shown in Table III.
For this mechanism only the vector component, $S_V$ gives non-zero contribution
($M^2_{inc} = S_V $). For the non-photonic mechanism we have non zero
contributions from both the vector and axial vector components, $S_V$ and
$S_A$, and the results are shown in Table IV ($M^2_{inc} = S_V + 3S_A$).

From Tables III and IV we see that, the main contribution to the incoherent
rate comes from the low-lying  excited states. High-lying excited states
contribute negligibly. This means that for a given A, a nuclear isotope
with many low-lying states in its spectrum 
is characterized by big incoherent matrix elements.  

For the doubly closed shell nucleus $^{208}Pb$, which is of current
experimental interest \cite{Ahmad,Honec}, the incoherent matrix elements are 
smaller than expected. 
A plausible explanation is that the spectrum of this nucleus 
presents a big gap (minimum energy needed to excite the first excited state) 
and only few excited states lie below $\approx 5 MeV$.

In general, the incoherent matrix elements do not show clear A and Z
dependence. Their magnitude depends on the spectrum of the individual
nuclear isotope.
  
In obtaining the results of Tables III and IV for $1^-$ states, we removed
the spurious center of mass contributions by explicitly calculating the
spurious state $|S>$ and removing its admixtures from the incoherent and
total rate. We have also calculated the overlaps $<1^-,m|S>$ (where $m$
counts the $1^-$ excited states) and found that most of the spuriousity 
lies in the lowest $1^-$ state being 88\% for $^{48}Ti$, 
63\% for $^{60}Ni$, 62\% for $^{72}Ge$, 57\% for $^{112}Cd$, 
87\% for $^{162}Yb$ and 77\% for $^{208}Pb$ \cite{Schwieger}.
We should stress that, for all nuclei studied, the spurious center of mass
contribution is less than 30\% of the incoherent matrix elements,
i.e., 1.0-1.5\% of the total $\mue$ conversion rate.

An additional point we should note is the effect of the ground state 
correlations on the $\mue$ matrix elements.
In QRPA this can be easily estimated by using a correlated quasi-particle RPA
vacuum instead of the uncorrelated one \cite{Row2}-\cite{McNe}.  In the
present work we have not performed additional calculations with a correlated
RPA vacuum. It is known, however, that the  matrix elements for $^{48}Ti$
obtained this way \cite{KVCF} are reduced by $\approx 30 \%$. The ground state
correlations tend to decrease the strengths of all $\mue$ conversion channels,
but do not affect the parameter $\eta$ (see sect. III.C). 

\bigskip
\centerline{\bf C. Comparison of coherent and incoherent processes}
\bigskip

As we have seen above, the $gs \ra gs $ channel is the most important one.
Therefore, a useful quantity for the $\mue$ conversion is the fraction of the
coherent matrix elements $M^2_{coh}$ divided by the total one 
$M^2_{tot}$, i.e. the ratio
 
\beq
\eta = M^2_{coh}/ M^2_{tot}
\label{eq:IIIC.1}
\eeq

In earlier calculations $\eta$ was estimated \cite{WeiFei} to be a
decreasing function of A with a value of $\eta \approx$ 83 \% in Cu region. By
using, however, the most appropriate QRPA(ii) results, we find that indeed the
coherent channel dominates throughout the periodic table (see Table V, for the
photonic and Table VI, for the non-photonic mechanism). In fact we see that the
values of $\eta$ obtained in the present calculations are a bit bigger than
those of Ref. \cite{Chia} obtained with a local density approximation and a lot
bigger than those of Ref. \cite{KV90}. We should stress, however, that the
exaggeration of the incoherent channels in shell model calculations of
Ref. \cite{KV90} is not a shortcoming of the method itself but the result of
ignoring the muon binding energy $\epsilon_b$ in calculating the nuclear form
factors. In fact, repeating the calculations of Ref. \cite{KV90} and taking
into account the effect of $\epsilon_b$ on the form factors of the coherent
process as well as on the mean excitation energy entering the total rate, we
find a value of $\eta \ge 75\% $.

\bigskip
\bigskip
\centerline{\bf IV. CONCLUSIONS}
\bigskip

In the present work we have studied in detail the dependence of $\mue$
conversion matrix elements on the nuclear parameters A and Z.  Our nuclear
matrix elements were obtained in the context of Quasi-particle Random Phase
Approximation (QRPA) which permits a relatively simple construction of all 
needed final states. So, there was no need to invoke closure approximation. 
The results obtained cover six nuclear systems from $^{48}Ti$ to $^{208}Pb$, 
which are of experimental interest. 
The most important conclusions stemming out of our detailed study are: 

i) The coherent mode dominates throughout the periodic table but it is more
pronounced in the heavy nucleus $^{208}Pb$ which is currently used at PSI
in the SINDRUM II experiment.
 
ii) The coherent and total rates as well as the ratio $\eta$ (coherent to total)
tend to increase as a function of the mass number A up to the $Pb$ region.
This encourages the use of heavier nuclear targets to look for lepton
flavour violation.

iii) In evaluating the nuclear matrix elements the muon binding energy
should not be ignored especially for heavy nuclear elements.

iv) The great part of the incoherent rate comes from the low lying
excitations.

The results obtained in the present work are in good agreement with those
obtained in the framework of the local density approximation  \cite{Chia} as
well as those of shell model calculations provided that all calculations
take into account the muon binding energy $\epsilon_b$.

\bigskip
\centerline{\bf ACKNOWLEDGMENTS }
\bigskip

This work has been partially supported (T.S.K.) by DFG  No FA67/19-1. The 
authors T.S.K. and J.D.V. would like to acknowledge support from the EU 
Human Capital and Mobility Program CHRX-CT 93-0323. 

\bigskip
\bigskip
\centerline{\bf APPENDIX }
\bigskip

{\bf A).}
The multipole expansion operators ${\hat T}^{(l,\sigma)J}$, resulting from
$\Omega_V$ and ${\bf \Omega}_A$  of Eq. (\ref{eq:IIA.1}), are written as

\beq
T_M^{(l,0)J}= \tilde g_V \del_{lJ} \, \sqrt{4\pi}\,
{\sum_{i=1}^A} (3+  \be \tau_{3i} )
 j_l (q r_i)Y_M^l ({\bf {\hat r}}_i)  \qquad
 \label{eq:A1}
\eeq

\noindent
for $\Omega_V$, the spin independent component, and 

\beq
T_M^{(l,1)J}=\tilde g_A \sqrt{\frac{4\pi}{3}}\,\,
{\sum_{i=1}^A} (\xi+ \be \tau_{3i}) j_l(q r_i)
\Big[ {Y^l({\bf {\hat r}}_i) {\bf \otimes} {\bf \sigma}_i} \Big] _M^J
\qquad 
\label{eq:A2}
\eeq

\noindent
for ${\bf \Omega }_A$, the spin dependent component. $j_l(qr)$ are the
spherical Bessel functions.

The quantities $W_{\la}^J \equiv W_{j_2j_1}^J$ of Eq. (\ref{eq:IIC.3}) 
contain the reduced matrix elements of the operators ${\hat T}^J$ between
the single particle proton or neutron states $j_1$ and $j_2$ as

 \beq
W_{j_1j_2}^J(\tau)\, =\, (\zeta + \tau \beta) \,\frac { {<j_1 \mid\mid 
{\hat T}^J\mid\mid j_2>} }{ {2 J }+1}
\label{eq:A3}
\eeq

\noindent
($\zeta =3$, for $\Omega_V$ and $\zeta=1/1.24$, for ${\bf \Omega}_A$).
The reduced matrix elements $<j_1||T^J||j_2>$ are given in Ref. \cite{KVCF}. 
The relevant radial matrix elements $<n_1l_1| j_l(qr) |n_2l_2>$, for harmonic 
oscillator basis often used, can be written in the elegant way

\beq
<n_1l_1| j_l(qr) |n_2l_2> \, = \,  e^{-\chi} 
\sum_{\kappa=0}^{\kappa_{max}} \, \varepsilon_{\kappa} \, \chi^{\kappa + l/2}
, \qquad  \chi = (qb)^2/4
\label{eq:A4}  
\eeq

\noindent
where

$$
\kappa_{max} = n_1 + n_2 +m, \qquad m=(l_1 + l_2 -l)/2
$$

\noindent
The coefficients $\varepsilon_{\kappa}(n_1l_1,n_2l_2,l)$, in general simple
numbers, are given by

\beq
\varepsilon_{\kappa} = \Big[ \frac{\pi \, n_1 ! n_2 !}{4 \,
\Gamma(n_1+l_1+\frac{3}{2}) \Gamma(n_2+l_2+\frac{3}{2})} 
\Big]^{\frac{1}{2}} \, \,
\sum_{\kappa_1 =\phi}^{n_1} \,\, \sum_{\kappa_2 =\sigma}^{n_2} \, n! \,
\Lambda_{\kappa_1}(n_1 l_1) \Lambda_{\kappa_2}(n_2 l_2)
\Lambda_{\kappa}(n l)
\label{eq:A5}  
\eeq

\noindent
where the $\Lambda_{\kappa}(n l)$ are defined in Ref. \cite{KVCF},
$n=\kappa_1 +\kappa_2 +m$ and

$$
\phi = \left\{ \begin{array}{ l@{\quad \quad} l}
0,  & \kappa - m -n_2 \le 0 \\ 
\kappa - m -n_2 , &  \kappa - m -n_2 > 0 \\ \end{array} \right. , \qquad
\sigma = \left\{ \begin{array}{ l@{\quad \quad} l}
0,  & \kappa - m -\kappa_1 \le 0 \\ 
\kappa - m -\kappa_1 , &  \kappa - m -\kappa_1 > 0 \\ \end{array} 
\right. 
$$

The advantage of Eq. (\ref{eq:A4}) is that, it permits the calculation of 
$\varepsilon_{\kappa}$, which are independent of the momentum q, once and for 
the whole model space used. Afterwards, the relevant reduced matrix elements 
are easily obtained for every value of the momentum transfer $q$.

\bigskip

{\bf B).} In the context of the quasi-particle RPA, the point-proton (-neutron)
nuclear form factors of Eq. (\ref{eq:IIB.5}) can be cast in the compact form

\beq
F_{\tau}(q^2) \,\,=\,\,
{1 \over {\tau}} e^{-(qb)^2/4}
\sum ^{N_{space}}_{\lambda =0} \theta^{\tau}_\lambda \,
(qb)^{2\lambda}, \qquad \tau = Z, \,\,N
 \label{eq:B1}
\eeq

\noindent
where $b$ is the harmonic oscillator parameter,
$N_{space}$ represents the maximum harmonic oscillator quanta included
in the model space used (see Table I), and
$\theta^{\tau}_\lambda$ the coefficients

\beq
 \theta^{\tau}_{\lambda} \,= \, \frac{\sqrt{\pi}}{2}
 {\sum _{(n,l)j, \, \,\,\,{ \lambda \ge l} } } \,
\Big( V^{\tau}_j \Big)^2
\,\, { {(2j+1) n! C^{\la -l}_{nl} } \over {\Gamma (n+l+{3 \over 2})} }
 \label{eq:B2}
\eeq

\noindent
where $\Big( V^{\tau}_j \Big)^2$ are the occupation probabilities for the
proton (neutron) single particle j-levels. The coefficients $C^m_{nl}$ are 
given in Ref. \cite{KV92}.


\newpage
\noindent
{\bf TABLE I.} Renormalization constants for proton ($g^p_{pair}$)
and neutron ($g^n_{pair}$) pairing interactions determined from the
experimental proton ($\Delta ^{exp}_{p}$) and neutron
($\Delta ^{exp}_{n}$) pairing gaps.

\vspace{0.3cm}

\begin{center}
\begin{tabular}{ccccccc}
\hline
\hline
& & & & & & \\
Nucleus & Configuration Space&
$b_{ho} (fm^{-1}$)& ${\Delta }^{exp}_{p} ( MeV )$ &
${\Delta }^{exp}_{n} ( MeV )$ & $g^p_{pair}$ & $g^n_{pair}$  \\
\hline
& & & & & & \\
 $^{48}_{22}Ti_{26}$ & 16 levels (no core) & 1.92 &
 1.896  & 1.564  &  1.082 & 1.002 \\
& & & & & & \\
 $^{60}_{28}Ni_{32}$ & 16 levels (no core) & 2.02 &
 1.718$^a$ & 1.395$^a$ & 1.033 & 0.901\\
& & & & & & \\
 $^{72}_{32}Ge_{40}$ & 16 levels (no core) & 2.07 &
 1.611 & 1.835 & 0.924 & 0.995 \\
& & & & & & \\
 $^{112}_{48}Cd_{64}$ & 16 levels (core $^{40}_{20}Ca_{20}$)& 2.21 &
 1.506 & 1.331 & 1.099& 0.950 \\
& & & & & & \\
 $^{162}_{70}Yb_{92}$ & 23 levels (core $^{40}_{20}Ca_{20}$) & 2.32 &
 1.170 & 1.104 & 0.894  & 0.951 \\
& & & & & & \\
 $^{208}_{82}Pb_{126}$ & 18 levels (core $^{100}_{50}Sn_{50}$) & 2.40 &
 0.807$^a$ & 0.611$^a$ &  0.861  & 1.042 \\
 \hline
\hline
 \end{tabular}
\end{center}

\vspace{0.4cm}
$^a$ {\it For the closed shell nuclei the parameters} $g^p_{pair}$ {\it and}
$g^n_{pair}$ {\it have been borrowed from the} $(N\pm 2,Z \mp 2)$ {\it nuclei i.e.
the experimental gaps} ({\it columns} 4 {\it and} 5) {\it for} $^{60}_{28}Ni_{32}$
{\it and} $^{208}_{82}Pb_{126}$, {\it are those of} $^{60}_{26}Fe_{34}$ {\it and}
$^{208}_{84}Po_{124}$, {\it respectively.}

\vskip0.8cm
\vspace{0.3cm}
\noindent
{\bf TABLE II.} Nuclear form factors for protons ($F_Z$) and neutrons ($F_N$)
calculated in the context of the shell model \cite{KV90} and quasi-particle RPA
cases: QRPA(i) and QRPA(ii) (see text). For comparison the experimental form 
factors \cite{Heisen,Vries} are also shown.

\vskip0.3cm
\begin{center}
\begin{tabular}{lcrrcccccc}
\hline
\hline
& & & & & & & & & \\
 Nucleus  & \multicolumn{3}{c}{ Shell Model } &
       \multicolumn{2}{c}{$  $ QRPA(i) $  $ } &
       \multicolumn{3}{c}{ QRPA(ii)} & Exper.\\
\hline
& & & & & & & & & \\
$(A,Z)$& $b_{ho}$ $(fm^{-1})$ & $F_Z $&$ F_N $&
$ $ $ F_Z $ $ $ &$ $ $ F_N $ $ $ &
 $\epsilon_b$ $(MeV)$ &$F_Z $&$ F_N$ & $F^{exp}_Z $ \\
\hline
& & & & & & & & & \\
$^{48}_{22}Ti_{26}$ & 1.906 & .543 & .528 & .528 & .506 &
1.250 & .537 & .514 & .532 \\
& & & & & & & & & \\
$^{60}_{28}Ni_{32}$ & 1.979 & .489 & .478 & .489 & .476 &
1.950 & .503 & .490 & .494 \\
& & & & & & & & & \\
$^{72}_{32}Ge_{40}$ & 2.040 & .470 & .448 & .456 & .435 &
2.150 & .472 & .451 & .443 \\
& & & & & & & &  & \\
$^{112}_{48}Cd_{64}$ & 2.202 & .356 & .318 & .349 & .312 &
 4.890 & .388 & .352 & .353 \\
& & & & & & & & & \\
$^{162}_{70}Yb_{92}$ & 2.335 & .261 & .208 & .252 & .218 &
 8.445 & .314 & .280 & .305\\
& & & & & & & & & \\
$^{208}_{82}Pb_{126}$ & 2.434 & .194 & .139 & .207 & .151 &
 10.475 & .271 & .214 & .242 \\
\hline
\hline
\end{tabular}
\end{center}

\newpage
\noindent
{\bf TABLE III.} Incoherent $\mu -e$ conversion matrix elements ($M^2_{inc}$) 
for the photonic mechanism. Only the vector component $(S_V)$ of the operator 
of Eq. (\ref{eq:IIA.1}) contributes. 

\vskip0.3cm 

\begin{center}
\begin{tabular}{cllllll}
\hline
\hline
        &         &        &         &       &          &         \\
$J^{\pi}$&$^{48}_{22}Ti$&$^{60}_{28}Ni$&$^{72}_{32}Ge$ 
&$^{112}_{48}Cd$&$^{162}_{70}Yb$&$^{208}_{82}Pb$\\
\hline
        &         &        &         &       &          &         \\
$ 0^+ $ & 1.946   & 1.160  &  2.552  &  2.088  &  4.305 &  2.512  \\
$ 2^+ $ & 0.242   & 0.738  &  1.396  &  2.669  &  6.384 &  2.342  \\
$ 4^+ $ & 0.004   & 0.005  &  0.015  &  0.021  &  0.063 &  0.056  \\
$ 6^+ $ &6 $10^{-6}$&6 $10^{-6}$&1 $10^{-5}$&6 $10^{-5}$  & 2 $10^{-4}$ 
 & 3 $10^{-4}$  \\
        &         &         &        &         &        &         \\
$ 1^- $ & 3.711   & 4.215   & 5.066  &  5.282  &  4.824 &  4.533  \\
$ 3^- $ & 0.037   & 0.081   & 0.152  &  0.249  &  0.542 &  0.476  \\
$ 5^- $ &2 $10^{-4}$&2 $10^{-4}$&5 $10^{-4}$& 0.001&0.005 &  0.005  \\
        &         &         &        &         &        &         \\
 $S_V$  & 5.940   &  6.199  &  9.181 & 10.309  & 16.123 &  9.924  \\
\hline
\hline
        &         &         &        &         &        &         \\
 $M^2_{inc}$ & 5.940  &  6.199 &  9.181 & 10.309 & 16.123 &  9.924 \\
\hline
\hline
\end{tabular}
\end{center}

\newpage
\noindent
{\bf TABLE IV.} Non-photonic mechanism. Incoherent $\mu -e$ conversion matrix 
elements ($M^2_{inc}$) for the vector $(S_V)$ and axial vector $(S_A)$ component
of the $\mue$ operator.

\vskip0.6cm
\begin{center}
\begin{tabular}{cllllll}
\hline
\hline
        &        &         &     &     &     &        \\
$J^{\pi}$&$^{48}_{22}Ti$&$^{60}_{28}Ni$&
$^{72}_{32}Ge$&$^{112}_{48}Cd$ &
$^{162}_{70}Yb$&$^{208}_{82}Pb$\\
\hline
        &         &        &         &         &        &        \\
$ 0^+ $ & 2.245   & 1.441  &  3.326  &  3.006  &  6.097 & 4.106  \\
$ 2^+ $ & 0.363   & 1.000  &  1.899  &  4.869  & 12.618 & 3.769  \\
$ 4^+ $ & 0.002   & 0.006  &  0.017  &  0.029  &  0.078 & 0.070  \\
$ 6^+ $ & 2 $10^{-6}$&7 $10^{-6}$&1 $10^{-6}$&8 $10^{-5}$ &2 $10^{-4}$
 & 4 $10^{-4}$ \\ 
        &         &         &        &         &        &         \\
$ 1^- $ & 4.010   & 6.164   & 6.676  &  7.243  &  6.796 &  5.537  \\
$ 3^- $ & 0.059   & 0.097   & 0.177  &  0.328  &  0.684 &  0.572  \\
$ 5^- $ & 1 $10^{-4}$ & 2 $10^{-4}$& 5 $10^{-4}$& 0.002 &  0.006 & 0.006 \\
\hline
        &         &         &        &         &   &   \\
 $S_V$  &   6.679 &  8.708  & 12.095 & 15.477  & 26.280  & 14.061 \\
\hline
\hline
        &         &        &         &         &       &        \\
$ 1^+ $ & 0.265   & 1.114  &  0.795  &  0.943  & 1.599 & 0.951  \\
$ 2^+ $ & 0.041   & 0.189  &  0.208  &  0.362  & 0.644 & 0.466  \\ 
$ 3^+ $ & 0.044   & 0.221  &  0.244  &  0.422  & 0.693 & 0.635  \\
$ 4^+ $ & 2 $10^{-4}$&0.001& 0.002   &  0.005  & 0.013 & 0.015  \\
$ 5^+ $ & 2 $10^{-4}$&0.002&  0.002  &  0.007  & 0.013 & 0.021  \\
$ 6^+ $ & 1 $10^{-7}$&1 $10^{-6}$&1 $10^{-7}$&1 $10^{-5}$ &4 $10^{-5}$
 &9 $10^{-5}$\\
        &         &         &        &         &   &  \\
$ 0^- $ & 0.770   & 1.498   & 1.394  &  1.840  & 1.925 & 1.539  \\
$ 1^- $ & 0.594   & 1.149   & 1.057  &  1.394  & 1.072 & 0.616  \\
$ 2^- $ & 0.673   & 0.800   & 0.880  &  1.035  & 1.137 & 1.170  \\
$ 3^- $ & 0.010   & 0.015   & 0.024  &  0.050  & 0.091 & 0.110  \\
$ 4^- $ & 0.009   & 0.019   & 0.031  &  0.057  & 0.164 & 0.129  \\
$ 5^- $ & 8 $10^{-6}$ & 2 $10^{-5}$&6 $10^{-5}$&2 $10^{-4}$&0.001 &
 0.001 \\
$ 6^- $ &9 $10^{-6}$ &7 $10^{-5}$&1 $10^{-4}$&3 $10^{-4}$&0.001 &
 0.001 \\
\hline
        &         &         &        &         &   &   \\
 $3 S_A$  &  2.405  & 5.009   & 4.637  & 6.115   & 7.305  & 6.226 \\
\hline
\hline
           &         &         &        &         &        &       \\
$M^2_{inc}$&  9.084  & 13.717  & 16.732 & 21.592  & 33.585 & 20.287 \\
\hline
\hline
\end{tabular}
\end{center}

\newpage
\noindent
{\bf TABLE V.} Coherent and total $\mu-e$ conversion rate matrix elements 
QRPA(ii) (see text) for the photonic mechanism in the neutrino mediated process. 
For comparison, we also show the ratio $\eta$ of Eq. (\ref{eq:IIIC.1}) given by
shell model \cite{KV90} obtained by ignoring the muon binding energy 
$\epsilon_b$. The results of QRPA(i) are similar to those of the shell model.

\vskip0.5cm
\begin{center}
\begin{tabular}{lccccc}
\hline
\hline
& & & & \\
Nucleus  & \multicolumn{2}{c}{ QRPA(ii) Matrix Elements }
         & \multicolumn{2}{c}{ $\eta \%$ } \\
\hline
& & & & \\
$(A,Z)$&$M^2_{gs \ra gs}$&$M^2_{tot}$& QRPA(ii)& Shell Model\\ 
\hline
& & & & \\
$^{48}_{22}Ti_{26}$ & 139.6 & 145.5  & 95.9  &       \\
& & & & \\
$^{60}_{28}Ni_{32}$ & 198.7 & 204.9  & 96.9  & 64.9  \\
& & & & \\
$^{72}_{32}Ge_{40}$ & 227.8 & 237.0  & 96.1  & 59.7  \\
& & & & \\
$^{112}_{48}Cd_{64}$& 346.7 & 357.0  & 97.1  &       \\
& & & & \\
$^{162}_{70}Yb_{92}$& 484.3 & 500.4  & 96.8  & 36.9  \\
& & & & \\
$^{208}_{82}Pb_{126}$&494.7 & 504.6  & 98.0  & 25.5  \\
\hline
\hline
\end{tabular}
\end{center}


\vskip0.8cm
\noindent
{\bf TABLE VI.} 
The same as in Table V for a non-photonic mechanism ($\beta = 5/6$)
in the neutrino mediated process. For comparison we have added the results 
for $\eta$ obtained with Local Density Approximation (L.D.A.) 
\cite{Chia} and shell model \cite{KV90}.  

\vskip0.5cm
\begin{center}
\begin{tabular}{lcccccc}
\hline
\hline
& & & & & \\
Nucleus &\multicolumn{2}{c}{ QRPA(ii) Matrix Elements }& 
&\multicolumn{2}{c}{ $\eta \%$ }  \\
\hline
& & & & & \\
(A,Z)&$M^2_{gs \ra gs}$ & $M^2_{tot}$ & QRPA(ii) & Shell Model & L.D.A. \\ 
\hline
& & & & & \\
$^{48}_{22}Ti_{26}$  & 375.2 & 384.3 & 97.6 &      & 91.0 \\ 
& & & & & \\
$^{60}_{28}Ni_{32}$  & 527.4 & 541.1 & 97.5 & 74.9 &      \\
& & & & & \\
$^{72}_{32}Ge_{40}$  & 639.5 & 656.2 & 97.4 & 70.3 &      \\
& & & & & \\
$^{112}_{48}Cd_{64}$ & 983.3 &1004.9 & 97.8 &      & 93.0 \\
& & & & & \\
$^{162}_{70}Yb_{92}$ &1341.2 &1374.8 & 97.6 & 40.1 &      \\
& & & & & \\
$^{208}_{82}Pb_{126}$&1405.2 &1425.5 & 98.5 & 28.2 & 94.0 \\
\hline
\hline
\end{tabular}
\end{center}

\newpage
\vspace*{10mm}
\centerline{\large \bf Figure Captions}

\vspace{15mm}
\noindent
{\bf FIGURE 1.} Typical diagrams entering the $\mue$ conversion:
The photonic (a), Z-exchange (b),(c) and box (d) diagrams. Only the specific 
mechanism involving intermediate neutrinos is exhibited here.
$\nu_e =\sum_j U_{ej} \nu_j$, $\,$ $\nu_{\mu} =\sum_j U_{\mu j} \nu_j$, 
where $\nu_j$ are the neutrino mass eigenstates and $U_{ej}$, $U_{\mu j}$
are the charge lepton current mixing matrix elements.  
Other mechanisms can also contribute (SUSY, $Z^{\prime}$, Higgs etc., 
see Ref. \cite{KLV94}).

\vspace{20mm}
\noindent
{\bf FIGURE 2.} Variation of the coherent $\mue$ conversion matrix elements 
$M^2_{coh}$ for specific mass A and charge Z (see text) for the photonic
mechanism (a) and the non-photonic mechanism (b). 
In QRPA(i) the muon binding energy $\epsilon_b$ was neglected, but it was
included in QRPA(ii). We see that $\epsilon_b$ strongly affects the matrix
elements for heavy nuclei. For comparison the results of Ref. \cite{KV90}
(shell model results) are also shown. For photonic and non-photonic diagrams
the coherent rate increases up to $Pb$ region where it starts to decrease.



\bigskip
\bigskip

\noindent
\begin{thebibliography}{99}
\bibitem{DOHM}C. Dohmen {\it et al.}, (SINDRUM II Collaboration),
Phys. Lett. {\bf B 317}, 631 (1993).
\bibitem{Ahmad}S. Ahmad {\it et. al.}, (TRIUMF Collaboration),
Phys. Rev. {\bf Lett. 59}, 970 (1987);
 Phys. Rev. {\bf D 38}, 2102 (1988).
\bibitem{SINDR}A. Badertscher {\it et al.}, (SINDRUM Collaboration),
J. of Phys. {\bf G 17}, S47 (1991);

A. van der Schaaf, Nucl. Phys. {\bf A 546}, 421c (1992); Prog. Part. Nucl. 
Phys.  {\bf 31}, 1 (1993). 
\bibitem{Honec}W. Honecker {\it et al.}, (SINDRUM II Collaboration),
Phys. Rev. {\bf Lett. 76}, 200 (1996).
\bibitem{WeiFei}S. Weinberg and G. Feinberg, Phys. Rev. {\bf Lett. 3}, 111
(1959); {\it ibid} E 244.
\bibitem{Shank}O. Shanker, Phys. Rev {\bf D 20}, 1608 (1979).
\bibitem{VER}J.D. Vergados, Phys. Reports {\bf 133}, 1 (1986).
\bibitem{Depom}P. Depommier and C. Leroy, Rep. Prog. Phys. {\bf 58}, 61 (1995).
\bibitem{KV90}T.S. Kosmas and J.D. Vergados, Nucl. Phys. {\bf A 510}, 641
 (1990).
\bibitem{KLV94}T.S. Kosmas, G.K. Leontaris and J.D. Vergados,
 Prog. Part. Nucl. Phys. {\bf 33}, 397 (1994);

T.S. Kosmas and J.D. Vergados, Phys. Reports {\bf 264}, 251 (1996).
\bibitem{INS}(MELC Collaboration), V.S. Abadjev {\it et al.},
preprint INS/Moscow (1992);

V.M. Lobashev, Private Communication (1996).
\bibitem{Brym}D.A. Bryman {\it et al.}, Phys. Rev {\bf Lett. 28}, 1469 (1972);
Phys. Rev. {\bf Lett. 55}, 465 (1985);

A. Badertscher {\it et al.}, Nucl. Phys. {\bf A 377}, 406
(1982).
\bibitem{Leonta}G.K. Leontaris, Study of the muonic and lepton numbers 
in modern gauge theories, Ph.D. thesis University of Ioannina, 1986, Greece.
\bibitem{MarSan}W.J. Marciano and A.I. Sanda, Phys. Rev. {\bf Lett. 38}, 
1512 (1977).
\bibitem{Berna}J. Bernabeu, E. Nardi and D. Tommasini, Nucl. Phys. {\bf B 409},
69 (1993). 
\bibitem{KV88}T.S. Kosmas and J.D. Vergados, Phys. Lett. {\bf B 215}, 460
(1988).
\bibitem{KV89} T.S. Kosmas and J.D. Vergados, Phys. Lett. {\bf B 217}, 19 
(1989).
\bibitem{Chia}H.C. Chiang, E. Oset, T.S. Kosmas, A. Faessler and
J.D. Vergados, Nucl. Phys. {\bf A 559}, 526 (1993).
\bibitem{KVCF}T.S. Kosmas, J.D. Vergados, O. Civitarese and A. Faessler,
Nucl. Phys. {\bf A 570}, 637 (1994).
\bibitem{Suho93}J. Suhonen, J. Phys. {\bf G 19}, 139 (1993).
\bibitem{Goul-Prim}B. Goulard, and H. Primakoff,
Phys. Rev. {\bf C 10}, 2034 (1974).
\bibitem{KV92} T.S. Kosmas and J.D. Vergados, Nucl. Phys.
 {\bf A 523}, 72 (1992).
\bibitem{KFSV94}T.S. Kosmas, A. Faessler, F. \v Simkovic and J.D. Vergados, 
Proc. 5th Hellenic Symp. on Nucl. Phys., Ioannina 1-2 Oct., 1993,
ed. X. Aslanoglou {\it et al.}, (University Press, 1994) p. 216.
\bibitem{Heisen}J. Heisenberg, R. Hofstadter, J.S. McCarthy and I. Sick,
Phys. Rev. {\bf Lett. 23}, 1402 (1969);

T.W. Donnelly and J.D. Walecka, Ann. Rev. Nucl. Sci. {\bf 25}, 329 (1975);

B. Frois and C.N. Papanicolas, Ann. Rev. Nucl. Part. Sci. {\bf 37}, 133 (1987).
\bibitem{Vries} H. de Vries, C.W. de Jager and C. de Vries, Atomic Data
 and Nuclear Data Tables, {\bf 36} 495 (1987).
\bibitem{Lala}G.A. Lalazissis and C.P. Panos, Phys. Rev. {\bf C 51}, 
1247 (1995).
\bibitem{Holin}K. Holinde, Phys. Reports {\bf 68}, 121 (1981).
\bibitem{Suho88} J. Suhonen, I. Taigel and A. Faessler, Nucl. Phys.
{\bf A 486}, 91 (1988).
\bibitem{CheSim} M.K. Cheoun, A. Bobyk, A. Faessler, F. \v Simkovic and
G. Teneva, Nucl. Phys. {\bf A 561}, 74 (1993); Nucl. Phys. {\bf A 564}, 
329 (1993).
\bibitem{Schwieger}J. Schwieger, T.S. Kosmas and A. Faessler, in preparation.
\bibitem{Duplain}D. Duplain, B. Goulard, and J. Joseph, Phys. Rev. 
{\bf C 12}, 28 (1975).
\bibitem{Row2} D.J. Rowe, Nuclear collective motion, (Methuen and CO. LTD., 
London, 1970).
\bibitem{Sande}E.A. Sanderson, Phys. Lett. {\bf 19}, 141 (1965);
 
J. Da Providencia, Phys. Lett. {\bf 21}, 668 (1966).
\bibitem{PElli}P.J. Ellis, Nucl. Phys. {\bf A 467}, 173 (1987).
\bibitem{McNe}J.A. McNeil, C.E. Price and J.R. Shepard,
Phys. Rev. {\bf C 42}, 2442 (1990).
\end{thebibliography}
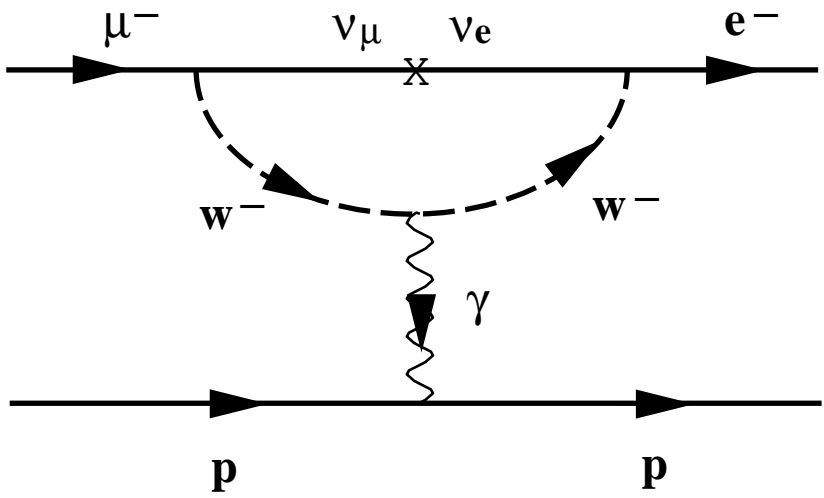
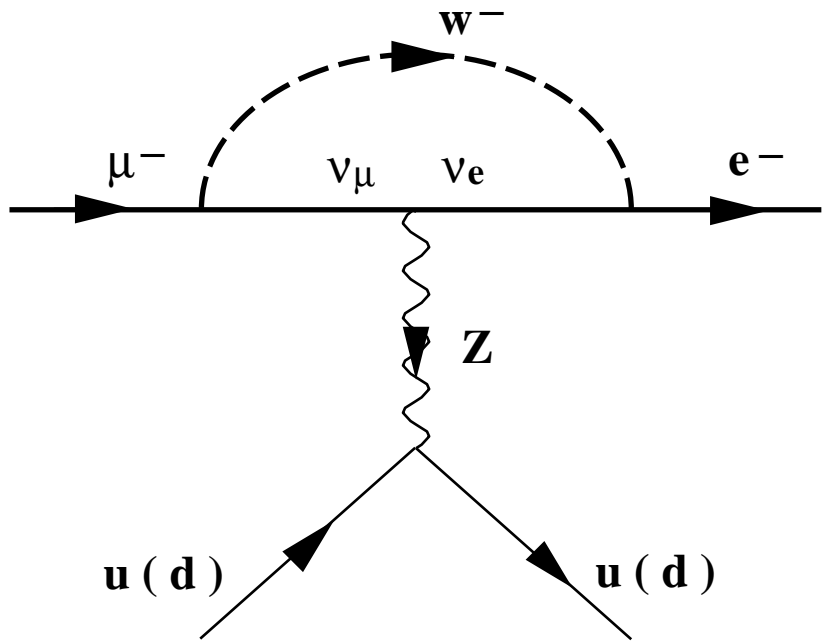
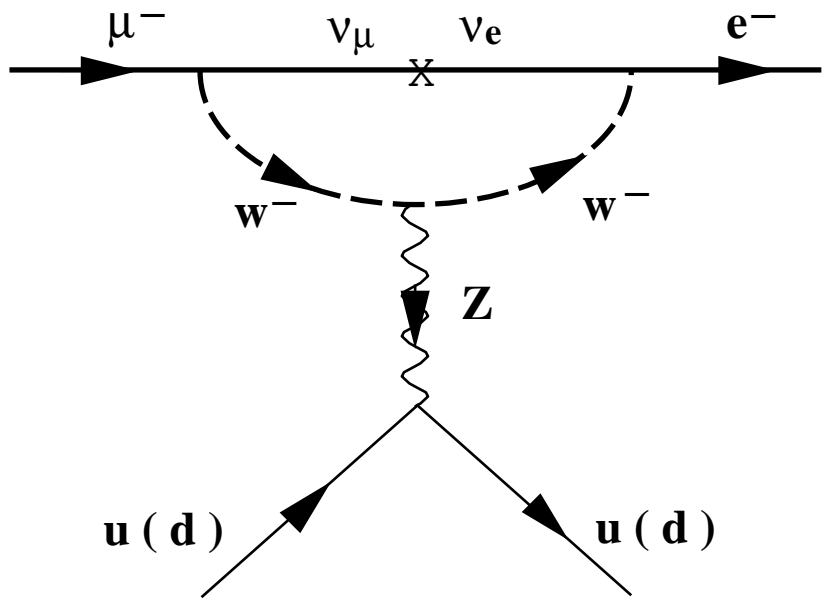
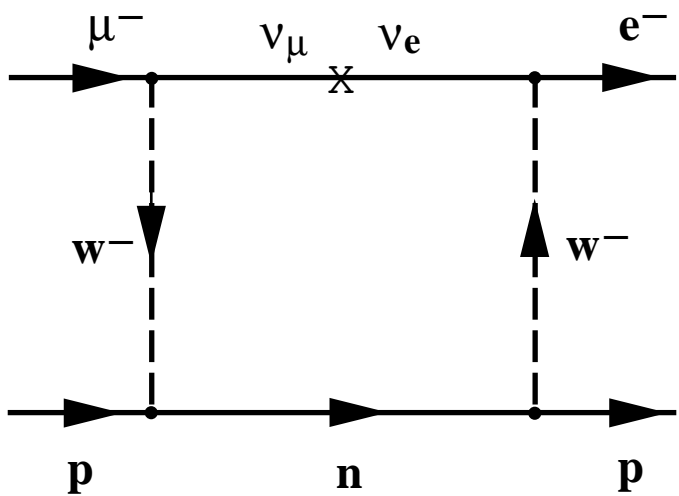
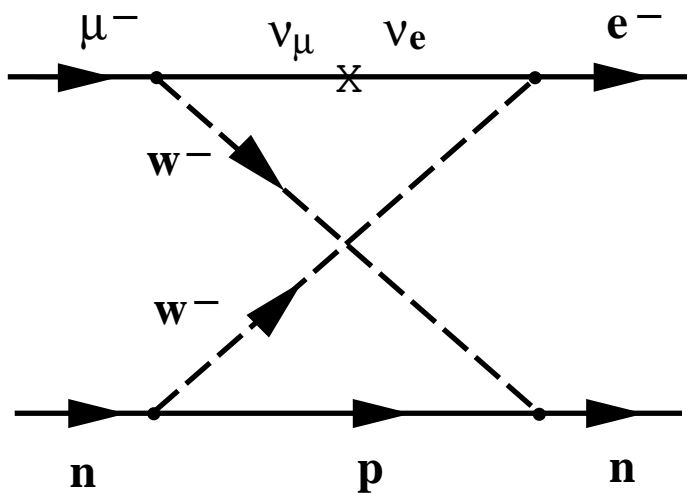
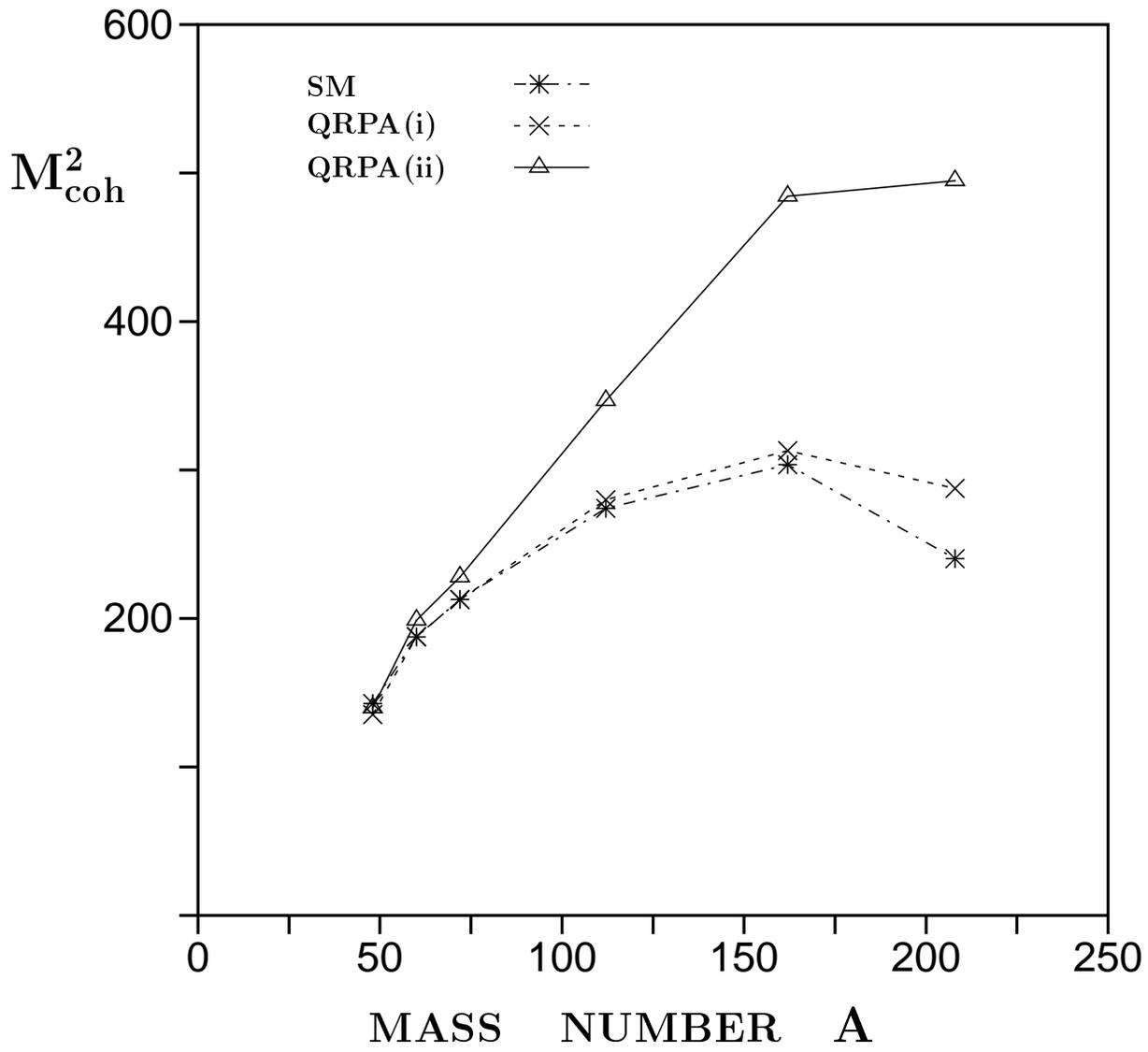
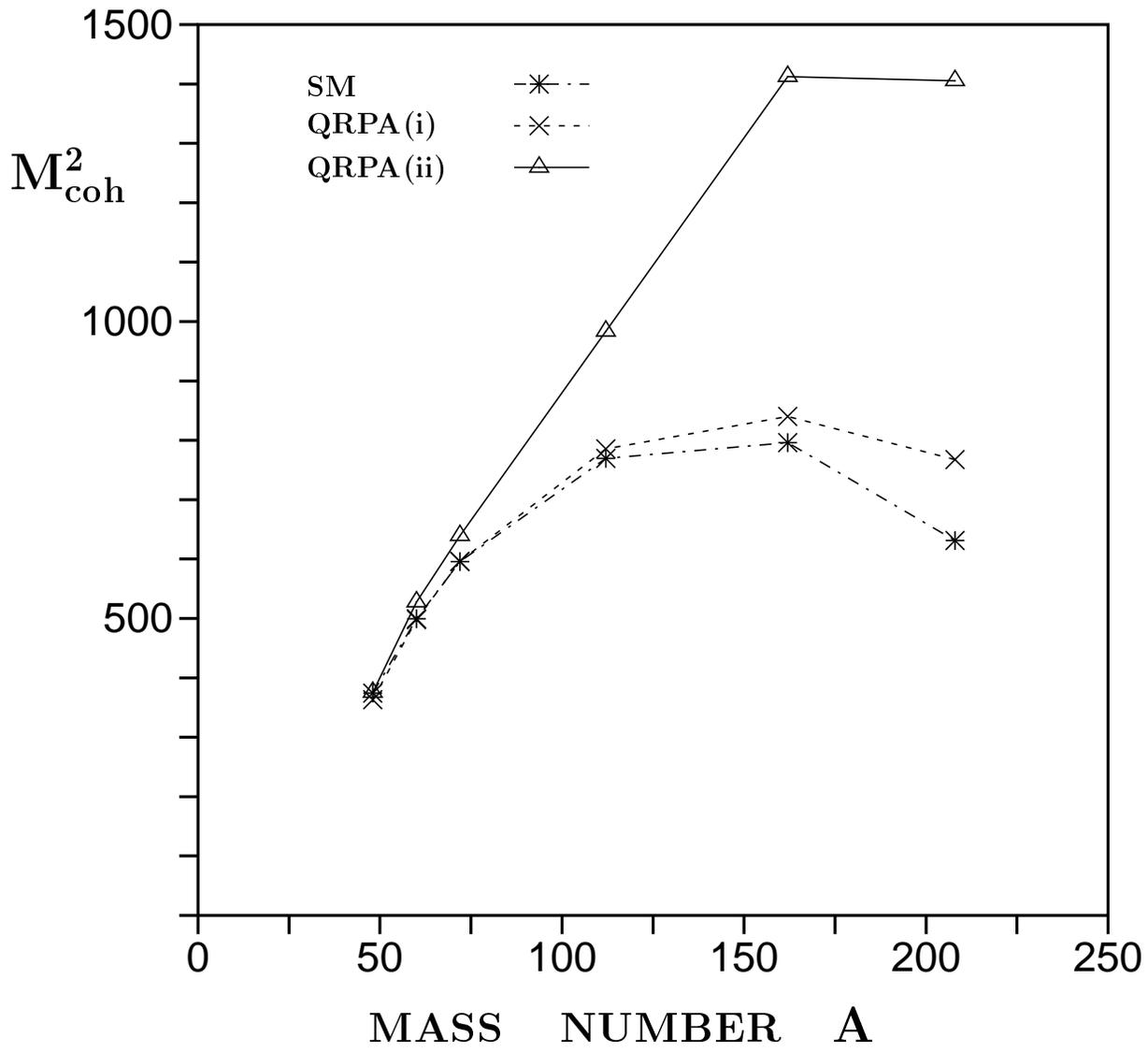
\end{document}